# ARE DARK MATTER AND DARK ENERGY THE RESIDUE OF THE EXPANSION-REACTION TO THE BIG BANG ?

**Harry I. Ringermacher**
General Electric Research Center, Schenectady, NY  12301 , USA

**Lawrence R. Mead**
Department of Physics and Astronomy
University of Southern Mississippi, Hattiesburg, MS  39406,  USA

**ABSTRACT**

We derive the phenomenological Milgrom square-law acceleration, describing the apparent behavior of dark matter, as the reaction to the Big Bang from a model based on the Lorentz-Dirac equation of motion traditionally describing radiation reaction in electromagnetism but proven applicable to "expansion reaction" in cosmology. The model is applied within the Robertson-Walker hypersphere, and suggests that the Hubble expansion exactly cancels the classical reaction imparted to matter following the Big Bang, leaving behind a residue proportional to the square of the acceleration. The model further suggests that the energy density associated with the reaction acceleration is precisely the critical density for flattening the universe thus providing a potential explanation of dark energy as well.  A test of this model is proposed.

*Subject headings*:  cosmology: theory --- cosmology: dark matter --- gravitational waves

## 1. INTRODUCTION

Dirac (1938) derived the relativistically covariant generalization of the Abraham-Lorentz equation describing the self-forces on a point electron arising from its emission of radiation and its reaction to an external force. This is the Lorentz-Dirac (LD) equation. Radiation reaction (RR) theory following the LD equation is nicely summarized and applied in Barut (1964).  Hobbs' (1968) and others have even developed corrections of the LD equation for a curved space. Regardless, the LD equation only concerns the dynamical effects of radiation emission on the particle path resulting from an "external force".  Indeed, it has been shown by Ringermacher (1979) that the precise covariant form of the LD equation for an external "Minkowski force" is demanded by the Frenet-Serret properties, alone, of a curve in a 4-space and is thus a universal, covariant result, independent of the nature of the physics, except for conservation of energy which defines the Minkowski force. Thus, the LD equation should hold as well for gravitation as it does for electrodynamics. Geometry/matter must substitute for radiation. Electrodynamics allowed the identification of 2 constant coefficients, the mass and classical electron radius as the length scale on the curve. Application to gravitation will result in a new length scale more appropriate to cosmology. In the present work we will examine the results of applying the LD equation to the dynamics of a co-moving cosmology. We take the space to be essentially flat, disregarding curvature couplings, and consider the effect of the "external expansion force" which we take to be conservative and therefore a Minkowski force, on mass points (future galaxies) via the traditional LD equation.

MOND or Modified Newtonian Dynamics, as described by Milgrom (1983), is a phenomenological equation that, to a great extent, summarizes the behavior of astrophysical objects ranging from galaxies to clusters of galaxies without the need for dark matter.  It does so by "modifying Newton's laws", which is unpleasant at best. Bekenstein (2004) has recently promoted MOND into a more rigorous relativistic Tensor-Vector-Scalar (TeVeS) theory, taken more seriously.  MOND requires that for accelerations less than a fundamental acceleration, $a_0$, the force is proportional to $a^2/a_0$ and is otherwise Newtonian.  The fundamental acceleration, $a_0$, points in the direction of the gravitational force and so is a deceleration whose magnitude is approximately $1 \times 10^{-10}$ m/s$^2$.  The LD equation includes the reaction force on the particle and its relativistic correction, which is proportional to the square of the acceleration – a truly unique force found in physics.  We will show that the effective expansion force itself resulting from the Big

Bang could be the source of the Milgrom "square-law" acceleration. Moreover, we will derive the formula that Milgrom correctly guessed for the value of the fundamental acceleration, $a_0$, that scales his square-law, along with the length scale for the cosmological LD equation.

## 2. EXPANSION-REACTION AND THE ROBERTSON-WALKER METRIC

Weinberg (1972) has suggested that "neutrinos and gravitons" should be taken seriously as cosmic mass sources. However, Rees (1971) appears to have been the first to suggest that gravitational waves generated with the big bang may be influencing current cosmology – even accounting for missing mass. He was, however, referring to wavelengths of order 1-10 Mpc, which would be isotropic and thus random, much like the cosmic microwave background, producing only averaged effects on galaxies. Here we are describing an effective force in the Hubble expansion, acting on a cosmological scale, resulting directly from coherent gravitational waves initiated during inflation. The "force" we are referring to is therefore the effective force of the expansion on matter we see today, namely galaxies. This expansion force spans the universe as long "coherent" gravitational waves and cannot average out randomly unlike Rees' shorter waves. By the Cosmological Principle the "coherence" must take the form of a central force about every point of space – consistent with the galactocentric MOND acceleration.

The standard form of the electrodynamic LD equation including radiation reaction forces is:

$$F^\mu = m\dot{v}^\mu - \frac{2}{3}e^2[\ddot{v}^\mu + v^\mu(\dot{v}^\alpha \dot{v}_\alpha)] \ , \tag{1}$$

$$v^\mu = \frac{dx^\mu}{d\tau} \qquad v^\mu v_\mu = 1$$

Its Minkowski Force generalization shown by Ringermacher (1979) to be precisely three terms at any point of a curve in a 4-space and independent of any physical radiation condition is:

$$F^\mu = \sigma_1 \dot{v}^\mu + \sigma_2[\ddot{v}^\mu + v^\mu(\dot{v}^\alpha \dot{v}_\alpha)] + \sigma_3[\ddot{v}^\mu + 3v^\mu(\dot{v}^\alpha \ddot{v}_\alpha)] \tag{2}$$

The third term in eq.2 plays no known role in electromagnetism and will be ignored for the purposes of the present exposition, although it could ultimately be relevant. The coefficients are not in general constant. Let us rewrite eq.2 in the form of a "non-Newtonian" acceleration, $\alpha^\mu$:

$$\alpha^\mu = \dot{v}^\mu - l[\ddot{v}^\mu + v^\mu(\dot{v}^\alpha \dot{v}_\alpha)] \tag{3}$$

$l$ is a length scale to be determined. We shall adapt the LD coordinates to the Friedmann-Robertson-Walker (FRW) model. The FRW metric is:

$$d\tau^2 = c^2 dt^2 - R^2(t)\left(\frac{dx^2}{1-kx^2} + x^2 d\theta^2 + x^2 \sin^2\theta d\varphi^2\right).$$

where $x = r/R_0$ is the cosmological co-moving radial coordinate. Also, for co-moving coordinates,

$$R(t) = a(t)R_0, \tag{4}$$

where $R_0$ is the Hubble radius and $a(t)$ is the normalized expansion parameter. The sign of the 3-space curvature, $K = k/R_0^2$, is $k = 0, +1, -1$. We will choose +1 for our immediate calculations and consider other cases later. The effects of spatial curvature on the dynamics of eq. (3) can be neglected leaving only the expansion parameter. The 3-D "hypersphere" undergoing Hubble expansion is described by:

$$\frac{\dot{R}}{R} = \frac{\dot{a}}{a} = H(t) \equiv H \tag{5}$$

The Hubble constant, $H_0 = H(t_0)$, is taken as 70 km/s/Mpc. Weinberg (1972) has shown, for positive curvature, $k = +1$, that the 3-space co-moving perimeter (proper length) on the hypersphere is given by:

$$L(t) = 4R(t)\int_0^1 \frac{dx}{\sqrt{1-kx^2}} = 2\pi R(t), \tag{6}$$

Thus, expansion of the hypersphere requires:

$$\dot{L}(t) = 2\pi \dot{R}(t) \tag{7}$$

This relation enables us to relate spacelike lengths to observable timelike lengths, which we will need later. Confining the motion to line of sight (constant $\theta, \varphi$) on the hypersphere, defining $\dot{v}^1 = \ddot{L}$ and taking the small velocity limit of eq. (3) yields

$$\bar{\alpha} = \ddot{L} - \frac{l}{c}\dddot{L} + \frac{l}{c^3}\dot{L}(a_N^2), \tag{8}$$

where we have defined the Newtonian acceleration $\ddot{L} = \bar{a}_N$. This is the cosmological LD equation for a galaxy attached to co-moving coordinates. Its dynamics will describe the galaxy's response to a Minkowski force in the form of the Hubble acceleration spanning the Hubble sphere. One recognizes, of course, that it is the space expanding with galaxies "standing still". Nevertheless, any observer will see accelerating galaxies and surmise that a "force" is acting. Thus we will use these concepts interchangeably. The force accelerates galaxies relative to us, resulting in a reaction acceleration and relativistic correction – on the space itself – in light of the above note. In order to elucidate this equation, consider the Hubble expansion in the FRW formalism. Our equation of state is $p = w\rho c^2$, where $\rho c^2$ is the energy density associated with the accelerated expansion arising from the cosmological constant. The Friedmann equations are:

$$\left(\frac{\dot{a}}{a}\right)^2 = \frac{8\pi G}{3}\rho - \frac{Kc^2}{a^2} \tag{9a}$$

$$\frac{\ddot{a}}{a} = \frac{-4\pi G\rho}{3}(1+3w) \tag{9b}$$

The energy equation, for convenience, is:
$$\dot{\rho} = -3H\rho(1+w). \tag{9c}$$

In order to evaluate the non-Newtonian acceleration, eq.(8), we will need to calculate the various time derivatives of the proper length, $L$. From eqs. (4), (5) and (7), we find:

$$\frac{\dot{L}}{L} = H \tag{10}$$

Differentiating eq. (9b) and using eq.(9c) together with eqs.(4), (5) and (7) yields:

$$\frac{\dddot{L}}{\ddot{L}} = -H(2+3w) \tag{11}$$

The general co-moving LD acceleration becomes, using eqs.(8), (10) and (11):

$$\bar{\alpha} = \left(a_N\left[1 + \frac{lH}{c}(2+3w)\right] + \frac{2\pi lHR}{c^3}(a_N^2)\right)\hat{L} \tag{12}$$

Choosing the present cosmological constant dominated universe, $w = -1$, yields the co-moving LD acceleration for the present accelerating epoch:

$$\bar{\alpha} = \left(a_N(1 - \frac{l}{c}H) + \frac{2\pi HRl}{c^3}(a_N^2)\right)\hat{L} \tag{13}$$

This equation is the non-Newtonian acceleration of a galaxy responding to the Hubble expansion acceleration in the FRW formalism as seen in the present epoch.

## 3. THE FUNDAMENTAL MOND ACCELERATION

We see immediately in eq. (13) that there is a term proportional to the square of the acceleration alongside a Newtonian term altered by the reaction effects depending on the length scale, $l$, and the epoch. When $l$ is vanishingly small, only the usual Newtonian acceleration remains. Although this model does not directly move us into a galaxy, it is clear from the symmetry argument discussed earlier that this force must act equally on a galaxy from all directions. Thus the force seen inside a galaxy must be equal and opposite and is therefore central and inward. Therefore we will take the view that eq. (13) also describes at least the dark matter dominated outer regions of the interior of a galaxy. These are the regions with negligible Newtonian acceleration. Since the MOND force is proportional to the square of the Newtonian acceleration in the limit of small accelerations, we require the first term of eq. (13) to vanish, yielding the Hubble length scale, $l$, for the present epoch:

$$l = \frac{c}{H_0} \equiv R_0 \qquad (14)$$

We see immediately that this scale is cosmic in scope and at the same time consistent with the Milgrom hypothesis. Furthermore, identifying the coefficient of $(a_N^2)$ of (13) with the inverse of the MOND fundamental acceleration, $1/a_0$, yields precisely Milgrom's conjecture and small acceleration law:

$$a_0 = \frac{c^2}{2\pi R_0} = \frac{cH_0}{2\pi} = 1.08 \times 10^{-10} ms^{-2} \qquad (15)$$

$$\vec{\alpha} = \left(\frac{a_N^2}{a_0}\right)\hat{L} \qquad (16)$$

The direction of the MOND acceleration is the same as the Newtonian acceleration. The precise value of $a_0$ depends on the choice of $k$ in eq. 6 and the value of the Hubble constant. For $k=0$, $a_0 = 1.7 \times 10^{-10} ms^{-2}$. Typical values used for fitting $a_0$, for example, Begeman, et al. (1991) lie between these two limits. The vanishing of the first term in (13) simply says the particle (galaxy) is essentially in free-fall, following a geodesic as expected. We note that this model, carried out to the present extent, does not produce the Milgrom "fitting formula" as used for galactic rotation curve fits. However, it does provide a model for the non-Newtonian $a_N^2/a_0$ component with the correct value of the "fundamental acceleration". In order to apply the model within the galaxy, one must "unfreeze" the comoving coordinates, which is out of the scope of the present work.

## 4. REALITY OF DARK MATTER

Recent data seems to suggest that dark matter has a reality of its own (Clowe, 2006). That is to say, it is not likely that a modification of Newtonian gravitation, centered about normal matter, can account for the observation. Since the present paper describes MOND as a dynamic, non-gravitational effect to begin with, it cannot address the issue of separation of baryonic and dark matter centers without a deeper underlying structure.

## 5. DISCUSSION AND TEST

In summary, we have derived the MOND square-law acceleration, including the correct value of the "fundamental acceleration" from the LD equations of motion by assuming the MOND behavior is the result of cosmic expansion interacting with a galaxy resulting in a galactic reaction. The nature of the derivation suggests the interpretation that the acceleration of the Hubble expansion at the present epoch exactly cancels the reaction acceleration ($\dot{v}$ term) imparted to a galaxy, leaving behind the reaction residue ($a^2$ term). This then is the source for an "extra force" playing the role of dark matter interior to galaxies. The residual acceleration is identified with the MOND phenomenology since the value of the fundamental scaling accelerations, $a_0$, are the same. We note in the same context that the repulsive effects arising from dark energy exterior to the galaxy induced the "dark matter force". Thus, in this model, dark energy and dark matter could be intrinsically related.

That an external "universal" force is involved can be readily appreciated from the following argument. If one takes the LD scale coefficient, $l = c/H_0$, which is the Hubble radius, $R_0$, and identifies it with the Schwarzschild radius, $2GM/c^2$, as though on a horizon (since the mass of the universe lies within $R_0$, and the "interior" RW solution must match the "exterior" Schwarzschild solution at $R_0$), one obtains the "Hubble mass", $M_0 = c^3/2GH_0$, which is the mass of the universe, or approximately $10^{56}$gm. From $M$ and $a_0$ one can define a truly "fundamental force",

$$F_0 = M_0 a_0 = \frac{c^4}{4\pi G} , \qquad (17)$$

composed of fundamental constants, and interpretable as the force of the Big Bang following inflation driving the mass of the universe on average at the fundamental MOND acceleration. The work done by this expansion is easily obtained from the force, eq.(17), acting over the distance on the hypersphere, $2\pi R_0$, from eq.(6). When distributed over the present Hubble volume $(4/3)\pi R_0^3$ this work appears as an energy density, $3H_0^2 c^2/8\pi G$, using eq.(14). This is precisely $\rho_c$, the critical energy density required to flatten the universe. Thus, this simple model suggests that the "Dark Energy" driving the present acceleration derives from the work of the Hubble expansion.

Once the scale, $l$, is set for the present epoch, we see that MOND is valid only for $w = -1$. Thus, one way to test this "Expansion-Reaction" model is to examine MOND for fits to earlier galaxies, particularly those with $z > 0.5$ corresponding approximately to the epoch at $w = -1/3$ when the universe's deceleration reversed and the current acceleration began about 6 Gy ago. This would introduce an additional linear term causing traditional MOND to overestimate the excess mass.

It is the intent of this paper to provide a simple and compelling argument demonstrating that expansion-reaction may have played a significant role in the Big Bang scenario leaving behind one more marker, the Milgrom acceleration, as evidence.